\documentclass[twocolumn,english,superscriptaddress,aps,prl]{revtex4-1}
\usepackage[latin9]{inputenc}
\usepackage{amsmath}
\usepackage{amssymb}
\usepackage{graphicx}
\usepackage{babel}
\begin{document}

\title{Two-dimensional Bose gases near resonance: universal three-body effects}

\author{Mohammad S. Mashayekhi}
\affiliation{Department of Physics and Astronomy, University of British Columbia, Vancouver V6T 1Z1, Canada}
\author{Jean-S\'ebastien Bernier}
\affiliation{Department of Physics and Astronomy, University of British Columbia, Vancouver V6T 1Z1, Canada}
\author{Dmitry Borzov}
\affiliation{Department of Physics and Astronomy, University of British Columbia, Vancouver V6T 1Z1, Canada}
\author{Jun-Liang Song}
\affiliation{Institute for Quantum Optics and Quantum Information, Austrian Academy of Sciences, A-6020 Innsbruck, Austria}
\author{Fei Zhou}
\affiliation{Department of Physics and Astronomy, University of British Columbia, Vancouver V6T 1Z1, Canada}

\begin{abstract}
We report in this Letter the results of our investigation of 2D Bose gases beyond the dilute limit emphasizing the 
role played by three-body scattering events. We demonstrate that a competition between three-body 
attractive interactions and two-body repulsive 
forces results in the chemical potential of 2D Bose gases to exhibit a maximum 
at a critical scattering length beyond which these quantum gases possess a negative compressibility. 
For larger scattering lengths, the increasingly prominent role played by three-body 
attractive interactions leads to an onset instability at a second critical value. 
The three-body effects studied here are universal, fully characterized by the effective 2D scattering 
length $a_{2D}$ (or the size of the 2D bound states) and are, in comparison to the 3D case, 
independent of three-body ultraviolet physics. We find, within our approach, the ratios 
of the contribution to the chemical potential due to three-body interactions to the one due to two-body to be 
%
$0.27$
near the maximum of the chemical potential and 
%
$0.73$
in the vicinity of the onset instability.
\end{abstract}

\maketitle

Two-dimensional quantum many-body systems have been, for many years, a subject 
of fascination for condensed matter and nuclear physicists alike. More recently,
this topic also caught the attention of the cold atom community
with the realization of quantum Bose gases confined to
two-dimensional geometries~\cite{Hadzibabic06,Schweikhard07,Clade09,Hung11}.
These experimental studies have so far explored these cold atom systems at temperatures close to 
the Berezinskii-Kosterlitz-Thouless phase transition~\cite{Berezinskii72,Kosterlitz73,Mermin66}. 
They highlighted the loss of long-range order due to the proliferation of vortices above 
the transition temperature, and the existence of two-dimensional quasi-condensates 
with algebraic long-range order and long wavelength thermal fluctuations 
below the transition. However, the fundamental properties of 2D Bose gases near absolute zero, 
where quantum effects are dominant, have yet to be addressed. In particular, on both
theoretical and experimental sides, very little work has been carried out to study 2D Bose gases 
near resonance. The main purpose of this Letter is to provide new light on the properties of 
2D Bose gases in this limit.

Compared to 3D Bose gases near resonance, which received more attention
in recent years~\cite{PappCornell2008, PollackHulet2009, NavonSalomon2011, WildJin2012}, 
2D gases possess important advantages. First, the ratio between 
elastic and inelastic collision cross sections can be significantly enhanced when atoms are confined 
to two-dimensional traps~\cite{Li09}. Second, in 2D, trimers and few-body structures 
are all universal as the absolute energy scale of the spectrum is uniquely set by 
the two-body binding energy and is independent of the short distance property of three-body 
interactions~\cite{Bruch79,Nielsen97,Hammer04,Blume05}. This is distinctly different from the 
physics of Efimov states in 3D as, in this case, the absolute energy scale 
is set by the ultraviolet physics of three-boson scatterings~\cite{Efimov70}.

These advantages are related to the dramatic suppression of the low energy effective interactions 
and phase shifts by coherent interference in 2D Bose gases. In fact, for an arbitrary repulsive 
interaction, the low energy two-body scattering phase shifts are logarithmically 
small indicating an asymptotically free limit. This aspect of scattering theory plays a critical role in 
the physics of 2D dilute Bose gases. Most previous works on 2D Bose gases considered systems where the
range of the repulsive interactions or the core size of the hardcore bosons, $a_0$, were much smaller than the
inter-particle distances~\cite{Schick71,Popov72,Fisher88}. Consequently, the results of these
studies are only applicable when $\frac{1}{\ln(na^2_0)}$ ($n$ is the density of bosons) is much 
smaller than unity, a limit corresponding to dilute gases in 2D. Here, 
we focus on the physics beyond the dilute limit to study 
2D Bose gases prepared on the upper branch and interacting via a resonating contact interaction. 
Such a setup can be achieved experimentally through a combination of Feshbach resonance and optical 
confinement~\cite{Petrov01,Bloch08,Pietila12}. 
Theoretically, to study 2D near-resonance Bose gases, we introduce 
a 2D effective scattering length $a_{2D}$.
This new tuning parameter 
is formally defined as the position of the node in the wave function for
two scattering particles and is also identified as the size of the two-body bound state. In general, 
$a_{2D}$ can be tuned to values larger than the averaged interatomic distance and can even be infinite.

Our study of 2D Bose gases at large scattering lengths unveils that near resonance the properties
of these gases are primarily dictated by the competition between three-body attractive interactions and 
two-body repulsive forces. We also show that the energetics of 2D Bose gases near resonance are universal
as they only depend on the parameter $na^2_{2D}$. 
Finally, we investigate the behavior of the
chemical potential for a wide range of scattering lengths. We find that the chemical 
potential first increases with $a_{2D}$ but very quickly reaches a maximum at 
%
$\frac{1}{ln(na^2_{2D})} = -0.135$
beyond which the Bose gas develops a negative compressibility.
Increasing $a_{2D}$ further brings about an onset instability at 
%
$\frac{1}{ln(na^2_{2D})} = -0.175$.
We identify both critical values to result from the important role played by three-body attractive 
interactions. Within our approach, we can estimate the contributions from three-body interactions
to the two-body ones to be around 
%
$0.27$ near the maximum of chemical potential
and 
%
$0.73$ in the vicinity of the onset instability.

To carry out this study of 2D Bose gases, we employ a method previously developed to understand the physics of
3D Bose gases near resonance~\cite{Borzov11,Zhou12}. In this approach, the chemical potential of non-condensed 
particles, $\mu$, and the density of condensed atoms, $n_0$, are first introduced as given parameters.
The Hamiltonian describing such a condensate interacting with non-condensed atoms through a short range 
interaction is
\begin{eqnarray}
H&=&\sum_{\bf k} (\epsilon_{\bf k} -\mu) b_{\bf k}^\dagger b_{\bf  k}
+ 2 U_0 n_0 \sum_{\bf k} b^\dagger_{\bf  k} b_{\bf  k}
\nonumber \\
&+&\frac{1}{2} U_0 n_0\sum_{\bf k}
b^\dagger_{\bf  k}
b^\dagger_{-\bf  k}+
\frac{1}{2}U_0 n_0 \sum_{\bf k} b_{\bf  k}b_{-\bf  k}
\nonumber \\
&+&\frac{U_0}{2\sqrt{S}}\sqrt{n_0}
\sum_{{\bf k'},{\bf q}} b^\dagger_{\bf q} b_{\bf  k'+\frac{\bf q}{2}}
b_{-\bf  k'+\frac{\bf q}{2}}+h.c.
\nonumber \\
&+&\frac{U_0}{2 S} \sum_{{\bf k}, {\bf k'},{\bf q}} b^\dagger_{\bf k+\frac{\bf q}{2}} 
b^\dagger_{-\bf  k+\frac{\bf q}{2}} b_{\bf  k'+\frac{\bf q}{2}}
b_{-\bf  k'+\frac{\bf q}{2}}+h.c.
\label{Hamitonian}
\end{eqnarray}
Here $\epsilon_{\bf k}=\hbar^2{\bf k}^2/2m$, the sum is over non-zero momentum states, 
$S$ is the total area, and $U_0$ is the strength of the bare short range interaction.
Later, we will evaluate $n_0$ and $\mu$ self-consistently as a function of the 
2D scattering length, $a_{2D}$, and of the total density $n$. 

Once the full system energy density $E(n_0,\mu)$ is known, one can calculate $\mu_c$, 
the chemical potential for the condensed atoms, and $n-n_0$, the density of non-condensed atoms.
This step is achieved using the following thermodynamic relations
\begin{eqnarray}
\label{SCE}
\mu_c &=& \frac{\partial{E(n_0,\mu)}}{\partial n_0},~~~ n = n_0 -\frac{\partial{E(n_0,\mu)}}{\partial \mu}; \\ \nonumber
\mu &=& \mu_c (n_0,\mu).
\end{eqnarray}
As hinted above, in the ground state, one requires $\mu_c$, the chemical potential for the condensed
atoms, to be equal to the chemical potential $\mu$. This equilibrium condition, first emphasized in 
Ref.~\cite{Hugenholtz59}, yields a self-consistent equation. The evaluation of $E(n_0, \mu)$ for a 
given $\mu$ and $n_0$ is usually carried out diagrammatically~\cite{Beliaev58,Hugenholtz59}. 
To capture the role of three-body interactions and to compare it with two-body contributions, 
we restrict ourselves to the virtual processes involving only two or three excited atoms.  
Truncating the Hilbert space accordingly, we can then sum up all connected diagrams contributing 
to the energy density. Within this truncation scheme, only the irreducible two- and three-body effective 
interaction potentials $g_{2,3}$ appear in the final expression for $E(n_0,\mu)$.
In order to implement the self-consistency condition and simplify the computation of $E(n_0,\mu)$, we
introduce for the non-condensed or virtual atoms an additional parameter $\eta=\Sigma-\mu$ where 
$\Sigma(n_0, \mu)$ is the self-energy. 
Physically, $\eta$ can be understood as an energy shift
due to the interaction between condensed and non-condensed atoms.
Using the same series of diagrams as in our study of 3D Bose 
gases near-resonance~\cite{Borzov11}, but carrying 
out the calculations in two spatial dimensions, we obtain for the energy density 
\begin{eqnarray}
\label{ED}
E(n_0,\mu) &=&\frac{1}{2}n_0^2 g_2(2\eta) +\frac{1}{3!}n_0^3 \text{Re}~g_3(3\eta) \nonumber \\
\text{with}~~g_2(2\eta) &=& \frac{\hbar^2}{m}\frac{4\pi}{\ln \frac{B_2}{2\eta} },
~~~ g_{3}(3\eta) = 6 g^2_2(2\eta) g_3^*(3\eta) \nonumber \\
\text{where}~~g_{3}^*(3\eta) &=& \frac{\hbar^2}{m}  
\int \frac{4q dq}{2\eta+q^2}\frac{G^{'}_{3}(-3\eta,q)}{\ln \frac{B_2}{3q^2/4+3\eta}}.
\end{eqnarray}
$g_{2,3}$ stand for, respectively, the {\it renormalized} two- and three-body interactions in 
a condensate. We will discuss this point in more details below. $G^{'}_{3}(-3\eta, p)$ represents the 
three-atom off-shell scattering amplitude (corresponding to the sum of all N-loop contributions 
with $N = 1, 2, 3, ...$). $G^{'}_3$ is the solution to the following integral equation, 
where $\hbar$ and $m$ were exceptionally set to unity to improve readability,
\begin{widetext}
\begin{eqnarray}
G^{'}_{3}(-3\eta,p)=\int \frac{4q dq}{\ln \frac{B_2}{3q^2/4+3\eta}} \frac{1}{\sqrt{(3\eta+p^2+q^2)^2-(pq)^2}}
\big(\frac{-1}{2\eta+q^2}-G^{'}_3(-3\eta, q)\big).
\label{G3}
\end{eqnarray}
\end{widetext}
Note that in Eqs.~\ref{ED} and \ref{G3}, $B_2 = \Lambda \exp\left(\frac{4\pi\hbar^2}{U_0 m}\right)$
where $\Lambda$ is an energy cutoff related to the effective 
interaction range, $R^{*}$, via $\Lambda = \frac{\hbar^2}{m R^{*2}}$. As $B_2 = \frac{\hbar^2}{ma^2_{2D}}$,
$g_{2,3}$ are uniquely determined by the parameter $\frac{n \hbar^2}{m B_2}$ or $n a^2_{2D}$.

For repulsive interactions (or positive $U_0$), $B_2$ is larger than $\Lambda$ and so
$a_{2D}$ is bounded from above by the interaction range $R^*$. When $U_0$ is infinite 
(hardcore potential), $a_{2D}$ is equal to the core size $a_0$. For attractive 
interactions (or negative $U_0$), the case we focus on here, 
$B_2$ is precisely the dimer binding energy, and $a_{2D}$ is the size of the bound state and 
can well exceed $R^*$. As a consequence, $n a^2_{2D}$, the fundamental tuning parameter for $E(n_0,\mu)$, 
can take values larger than unity. The gas can hence be tuned away from the dilute 
limit~\cite{quasi2d}.

Before detailing our results, we need to mention, as a matter of completeness, that as $E(n_0,\mu)$ explicitly 
depends on $\Sigma(n_0, \mu)$, $n_0$ and $\mu$, Eqs.~\ref{SCE} and \ref{ED} are supplemented 
by the relation
\begin{eqnarray}
\Sigma(n_0,\mu)=\mu_c(n_0,\mu) +\frac{\partial \mu_c}{\partial \ln n_0},
\label{HP} 
\end{eqnarray}
an extension of Hugenholtz-Pines theorem~\cite{Hugenholtz59}.

\begin{figure}
\includegraphics[width=0.95\columnwidth]{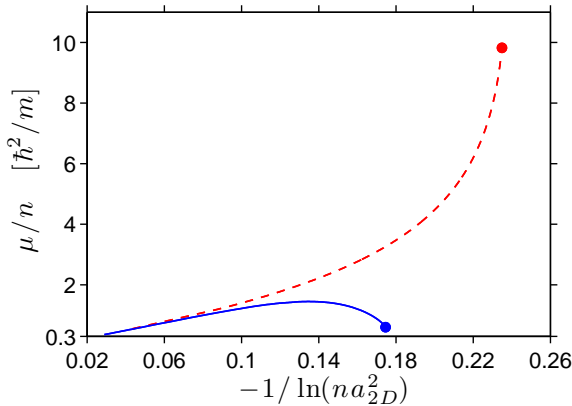}
\caption{(Color online) The chemical potential, in units of $\frac{\hbar^2n}{m}$, 
as a function of $n a_{2D}^2$. The dashed (red) line is the solution of 
the self-consistent equation when only two-body interactions are included. 
The full (blue) line is the solution when both two- and three-body interactions 
are included. This figure highlights that the behavior of the chemical 
potential is drastically altered by three-body physics.}
\label{fig:fig1}
\end{figure}

3D counterparts to Eqs.~\ref{SCE}, \ref{ED} and \ref{HP} were used in Ref.~\cite{Borzov11} 
to obtain the chemical potential of 3D Bose gases near resonance. These self-consistent equations 
provided highly precise estimates for the chemical potential in the dilute limit. Near 
resonance, this approach predicted a maximum in the chemical potential and an accompanied onset instability.
These features were fully consistent with the conclusions drawn from a renormalization group equation 
approach~\cite{Zhou12}. This first study concluded that in 3D the dominating contribution 
to the chemical potential came from irreducible two-body interactions; for cold atoms, the three-body 
contribution was negligible. For 2D Bose gases, the story is very different:
three-body interactions play here a much more important role as can be seen on Fig.~\ref{fig:fig1}.

To analyze the contribution coming from the three-body effect, we first solve Eqs.~\ref{SCE} and \ref{ED} excluding 
the contribution of $g_3$, and obtain the chemical potential solely due to two-body interactions
(see Fig.~(\ref{fig:fig1}) dashed red line). Here, $g_{2}$ is defined as the effective two-body interaction 
renormalized by scattering events off condensed atoms and includes a subset of $N$-body interactions defined in 
the vacuum~\cite{1loop}. Neglecting $g_3$ interactions, the self-consistent equations take the simple form 
\begin{eqnarray}
\tilde{\mu} &=& \frac{4\pi}{\ln \frac{1}{2\alpha \tilde{\mu}}}+\frac{8\pi^2}{\tilde{\mu} \ln^3 \frac{1}{2 \alpha \tilde{\mu}}},~~
\frac{1}{\tilde{n}_0} = 1+\frac{2\pi}{\tilde{\mu}}\frac{1}{\ln^2 \frac{1}{2\alpha \tilde{\mu}}}
\label{G2SCE}
\end{eqnarray}
where $\tilde{\mu}=\frac{m\mu}{\hbar^2n_0}$, $\tilde{n}_0 =\frac{n_0}{n}$ and $\alpha = n_0 a^2_{2D}$~\cite{Approx}. 
The solution of Eq.~\ref{G2SCE} in the limit of small $\alpha$ is
%
%
\begin{eqnarray}
\mu=\frac{n}{m}\frac{4\pi\hbar^2}{\ln\frac{1}{\alpha}}\left(1 - \frac{1}{\ln\frac{1}{\alpha}}
[\ln |\ln \alpha| - \ln 4\pi +C]+...\right)
\end{eqnarray}
%
where $C = \ln \frac{1}{2}$ 
within this self-consistent approach.
This solution, valid in the dilute limit, agrees well with previous studies~\cite{Schick71,Popov72,Beane10}.
Another solution with $\mu$ approaching $\frac{\hbar^2}{ma_{2D}^2}$ exists in this limit but is unstable.
As $\alpha$ or $na^2_{2D}$ is increased, the dilute gas solution approaches this higher energy unstable solution,
and at the critical value 
%
$na^2_{2D}= 1.42 \times 10^{-2}$
the two solutions coalesce into one. 
Beyond this point, no real solution to Eq.~\ref{G2SCE} exists revealing the presence of an instability.
The basic structure sketched here, when three-body contributions are neglected, is qualitatively 
the same as that of 3D Bose gases: $\mu$ is maximum when an onset instability
sets in, and for larger $na_{2D}^2$ develops an imaginary part implying the formation of molecules.

\begin{figure}
\includegraphics[width=0.95\columnwidth]{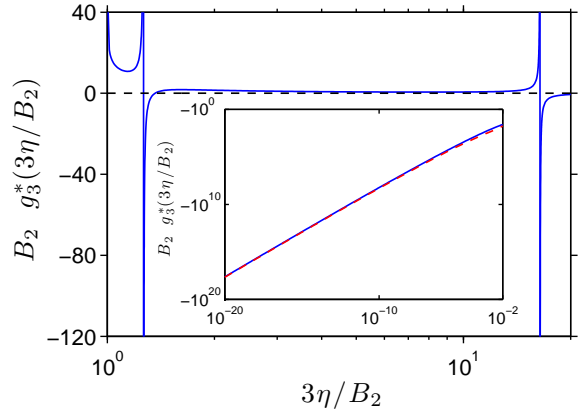}
\caption{(Color online)
Three-body interaction $g^*_3$ (defined in Eq.\ref{ED}) as a function of the energy shift $\eta=\Sigma-\mu$;
$\eta$ is determined self-consistently together with $\mu$. 
Inset: full and two-loop behavior of $g^*_3$ for small $\eta$ values (respectively, full (blue) 
and dashed (red) lines). For $\frac{3\eta}{B_2} < 1$, the numerical integration over the momentum was done from
$0$ to $50 \frac{\sqrt{B_2 m}}{\hbar}$.}
\label{fig:fig2}
\end{figure}

We now turn our attention to the contribution  of $g_3(3\eta)$. 
$g_3(3\eta)$ is obtained by first numerically solving Eq.~\ref{G3} 
for $G^{'}_3(-3\eta, p)$ and then by carrying out the integral 
involving $G^{'}_3(-3\eta,q)$ in Eq.~\ref{ED}. The result of this procedure 
is shown in Fig.~\ref{fig:fig2} where we plot $g_3^*(3\eta)$. We chose to
plot $g_3^*(3\eta)$ and not $g_3(3\eta)$ as the former is not cluttered by
trivial effects due to $g_2^2(2\eta)$. We identify two kinds of resonant
scattering processes defining the basic structure of $g_3(3\eta)$. 
The first one is a three-body resonance between 
three condensed atoms with zero energy and a dimer plus a non-condensed atom with total 
energy $3\eta -B_2$. Here $3\eta$ is the mean-field energy shift due to the exchange 
interaction between the non-condensed atom-dimer structure and the condensate.
This leads to the first peak (from left to right) at $3\eta = B_2$.
The second process is a three-body resonance between three condensed atoms and a trimer 
with either binding energy $B_3^{(1)}$ or $B_3^{(2)}$ (or total energies $3\eta - B_3^{(1)}$ 
or $3\eta -B_3^{(2)}$). This process produces the second and third peaks at $3\eta=B_3^{(1,2)}$. 
We find numerically that
%
$B_3^{(1)} = 1.296 B_2$ and
%
$B_3^{(2)} = 16.643 B_2$.
These energies are fully consistent with the results of two previous few-body 
studies~\cite{Bruch79,Hammer04}. Unlike in 3D where a logarithmically large number of Efimov states 
exist, in 2D there are only two trimer states. Remarkably, their energies are 
uniquely determined by $B_2$ without involving an additional three-body parameter,
a fascinating feature emphasized in Refs.~\cite{Bruch79,Hammer04}.

\begin{figure}
\includegraphics[width=\columnwidth]{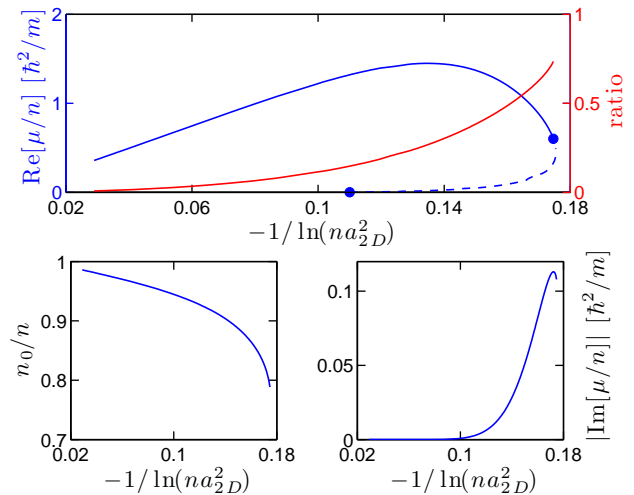}
\caption{(Color online) Top panel: ratio between the contributions of three-body 
and two-body interactions as a function of $na^2_{2D}$ (full red line),  
chemical potential for 2D Bose gases (full blue line). An additional 
metastable solution (dashed blue line) also exists 
when $g_3$ is included. The maximum value of $\mu$ is 
%
$1.45 \frac{\hbar^2n}{m}$
and occurs at 
%
$na_{2D}^2 = 0.604 \times 10^{-3}$.
Bottom left panel: condensation fraction $n_0/n$ 
as a function of $na^2_{2D}$.
Bottom right panel: imaginary part of the chemical potential when taking into account the 
contribution of all three-body recombination processes.
Note that $|\text{Im}~\mu| \ll \text{Re}~\mu $ for all considered $na^2_{2D}$, indicating
the quasi-static nature of the Bose gases. Hence, three-body recombination 
plays very little role in our energetic analysis and can be safely neglected for the 
range of parameters considered.}
\label{fig:fig3}
\end{figure}

The effect of three-body scatterings on the quantum gas is mainly determined 
by the property of $g_3$ when $\eta$ is relatively small. 
We checked numerically that in the limit of very small $\eta$, $g_3$ can be well fitted 
by an attractive interaction of the scaling form 
$\frac{\hbar^4}{2m^2\eta}\frac{1}{\ln^2\frac{B_2}{2\eta}\ln^2\frac{B_2}{3\eta}}$, 
capturing the dominant two-loop contribution~\cite{Nloop} (see Fig.~\ref{fig:fig2}).
Including the contribution due to three-body physics in the evaluation 
of the chemical potential results in two main effects. First, due to the attractive tail 
of $g_3$ in the small $\eta$ limit, as shown in Fig.~\ref{fig:fig3}, the instability 
is shifted away from 
%
$na^2_{2D}= 1.42 \times 10^{-2}$
and occurs at a much smaller value of 
%
$na_{2D}^2= 3.26 \times 10^{-3}$.
At this new instability point, the chemical potential is dramatically reduced, from 
%
$9.82 \frac{\hbar^2n}{m}$
to
%
$0.601 \frac{\hbar^2n}{m}$
when $g_3$ is included. In other words, the three-body effective interaction further 
destabilizes the quantum gas. The second and equally important effect is that the inclusion of 
three-body interactions results in the appearance of a maximum in the chemical potential at 
%
$na^2_{2D}= 0.604 \times 10^{-3}$
before the onset instability occurs. The maximum value of the chemical potential is 
%
$\mu_{max} = 1.45 \frac{\hbar^2n}{m}$
and the condensation fraction at the maximum is 
%
$91\%$.

Between the maximum and instability points, 
the quantum gas exhibits a negative compressibility and can potentially collapse into 
a high density phase. Although the fate of the Bose gases with negative compressibilities 
and the details of the corresponding dynamics are beyond the scope of our investigation,
we speculate that in this regime a quantum gas eventually evolves into the droplet matter 
discussed in Ref.~\cite{Hammer04}.
In 3D, the instability originated from a shift of the dimers due to scatterings off condensates 
and was a precursor of the sign change of the effective two-body interaction 
$g_2$~\cite{Zhou12}. For 2D Bose gases, the situation is completely different.
Here, the instability is a consequence of the competition between the repulsive two-body 
interaction (positive $g_2$) and the attractive three-body interaction (negative $g_3$) 
in the low energy limit.
For a 2D Fermi gas, the Pauli blocking effect was recently demonstrated to lead to an 
instability at a finite scattering length~\cite{Pietila12}.
 
We also plot in Fig.~\ref{fig:fig3} the relative weight of the three-body to 
two-body contributions to the chemical potential. As anticipated, the three-body contribution is negligible 
in the dilute limit when $na^2_{2D} \ll 1$ but quickly becomes important as $na^2_{2D}$ is 
increased. The prominent role played by three-body scattering leads to a maximum in the 
chemical potential before the instability point. At this maximum,
the ratio between the three-body and two-body contributions reaches 
%
$0.27$.
The shift of the instability is also caused by the attractive 
three-body interactions.

In conclusion, we demonstrated that the properties of 2D Bose gases 
at large scattering lengths 
or near resonance are dictated by three-body effects. 
We showed that the contributions of trimer states are universal as they only depend 
on the effective two-body scattering length $a_{2D}$ and not on the short distance properties of 
three-body interactions; an aspect unique to 2D Bose gases. Our results also suggest 
the existence of strong correlations in the three-atom channel near 
resonance. This feature remains to be probed experimentally.


F.Z. would like to thank Chen Chin for useful discussions.
This work is supported by NSERC (Canada), CIFAR, and the Izaak Wlaton Killam 
Memorial Fund for Advanced Studies. 



\begin{thebibliography}{References}
\bibitem{quasi2d}For a quasi two-dimensional cold gas near Feshbach resonance, $a_{2D}$ is a 
function of $l_0$, the confinement radius along the perpendicular direction, and of the free space 
scattering length $a_{3D}$. For shallow 2D bound states, 
$a_{2D}=\sqrt{\frac{\pi}{0.915}} l_0 \exp\left(-\sqrt{\frac{\pi}{2}} \frac{l_0}{a_{3D}}\right)$
~\cite{Petrov01,Bloch08,Pietila12}.
\bibitem{1loop}In the dilute limit, $g_2$ reproduces the most dominating contribution; the residue 
effects are from the irreducible $N = 4,6,...$-body interactions, and are parametrically smaller than the contributions 
from $g_2$, i.e. smaller by a factor of $\frac{1}{ln(na^2_{2D})}$. 
See also similar discussions on 3D cases in Ref.~\cite{Borzov11}.
\bibitem{Approx}In obtaining this equation, we take into account Eq.~\ref{HP} and set 
$\partial \eta/\partial \mu=1$, $\partial \eta/\partial n_0=g_2$ and $\eta=\mu$ to simplify the 
structure.
\bibitem{Nloop}A diagrammatic calculation similar to the one presented in Ref.~\cite{Borzov11} suggests that 
the leading $N$-loop contributions to $g_3$ for small values of $\eta/B_2$ are
$g_3^\text{(N-loop)}= C_N \frac{\hbar^4}{2m^2\eta}\frac{1}{\ln^{2} \frac{B_2}{2\eta} \ln^N \frac{B_2}{3\eta}}$
with $N=2,3,4...$; the prefactor $C_N$ can be computed numerically. For the most dominating two-loop 
contribution, 
%
$C_2 = -6.3 \times 10^{3}$.
\end{thebibliography}
\end{document}